\documentclass[]{spie}  

\usepackage{amsmath,amsfonts,amssymb}
\usepackage{graphicx}
\usepackage[colorlinks=true, allcolors=blue]{hyperref}
\usepackage{color}

\title{Quantum computational imaging and sensing}

\author[a]{Mohan Sarovar}
\affil[a]{Quantum Algorithms and Applications Collaboratory, Sandia National Laboratories, Livermore, CA 94550, USA}

\authorinfo{E-mail: mnsarov@sandia.gov}

\begin{document} 
\maketitle

\begin{abstract}
We present a new framework for imaging and sensing based on utilizing a quantum computer to coherently process quantum information in an electromagnetic field. We describe the framework, its potential to provide improvements in imaging and sensing performance and present an example application, the design of coherent receivers for optical communication. Finally, we go over the improvements in quantum technologies required to fully realize quantum computational imaging and sensing.
\end{abstract}

\keywords{Quantum computing, quantum sensing, quantum imaging}

\section{INTRODUCTION}
\label{sec:intro}  

\begin{figure}[h]
\centering
\includegraphics[width=\columnwidth]{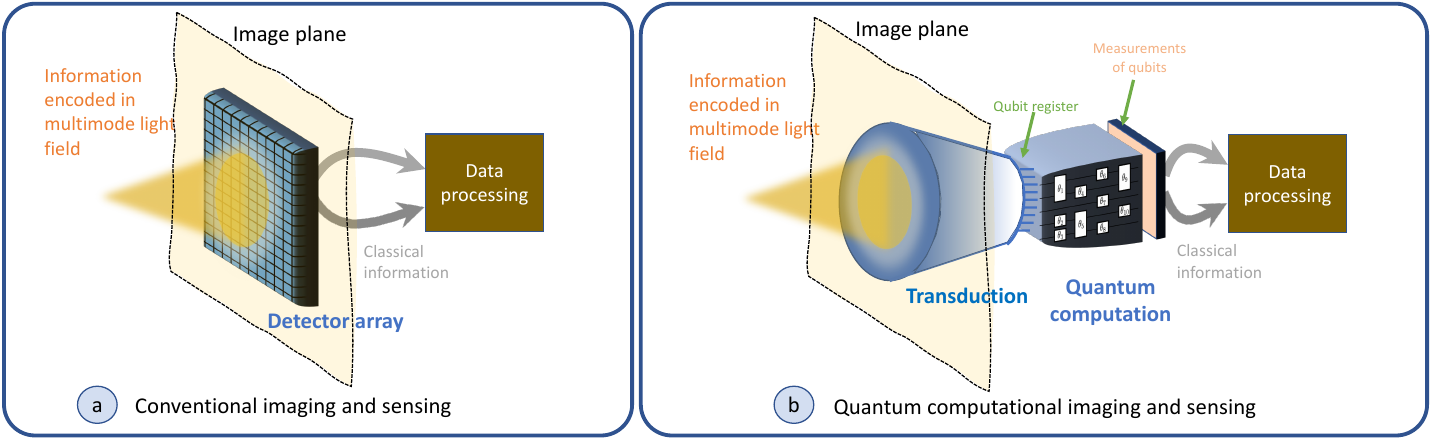}
\caption{\textbf{(a)} Conventional imaging and sensing, where quantum information in the electromagnetic field is immediately transformed to classical information by intensity detection at the image plane. \textbf{(b)} Schematic of the quantum computation imaging and sending (QCIS) concept, where relevant quantum information is transduced into the state of a register of qubits, on which a quantum computation is performed before conversion to classical information. \label{fig:qcis}}
\end{figure}

Recent progress in quantum technologies and quantum complexity theory can be combined to motivate a powerful new way to probe the physical world, whereby quantum states of the electromagnetic (EM) field are transduced into quantum computers before a quantum computation is performed to make inferences relevant to sensing, detection and imaging tasks, see Fig. \ref{fig:qcis}. This concept, which we term \emph{quantum computational imaging and sensing (QCIS)}, relies on two key quantum technologies: (i) universal quantum computers that are capable of transforming quantum states coherently with minimal loss and decoherence, and (ii) quantum transduction, which transfers quantum states between different carriers at varying energy scales. Transduction has mostly been considered in the context of quantum networking where it is the technique by which quantum states are transferred from stationary qubits to flying qubits such as photons \cite{Lauk_2020}. In the QCIS context, quantum transduction is a way to get \emph{unknown quantum states} of the EM field \emph{into} a quantum computer.

Coupled with developments in the two technologies described above, there have also recently been several foundational results in quantum complexity and quantum computer science that have established exponential separations for learning quantum states with and without quantum computers \cite{Chen_Cotler_Huang_Li_2021,Aharonov_Cotler_Qi_2022,Huang_2022}. Coarsely, these results show that one can learn properties of quantum states with exponentially fewer copies if these states can be stored in quantum memory and operated on coherently by a quantum computer. These theoretical results have some very practical repercussions, principal among these being a roadmap to achieve tremendous improvements in weak-field sensing and imaging. For example, the separations mentioned above imply that one could learn properties of an image in a field of view with exponentially less light/signal than currently required, \emph{if} the relevant quantum states of an EM field can be transduced with high fidelity into a register of qubits on which a quantum computation is performed. Provable exponential quantum advantages are rare, and realizing the advantages derived in Refs. \cite{Chen_Cotler_Huang_Li_2021,Aharonov_Cotler_Qi_2022,Huang_2022} in practical imaging and sensing settings is the primary motivation behind QCIS.

Abstractly, the QCIS framework uses a quantum computer to implement or approximate the most general measurement allowed by quantum mechanics on the EM field. This perspective also reveals the power of the framework: it enables the implementation of customized measurements that maximize information gain, achieve sensing tasks inaccessible to conventional classical sensors, and reduce the data burden at the collection point. This also suggests that QCS is a way to achieve a quantum advantage with noisy intermediate-scale quantum (NISQ) devices \cite{Preskill_2018} that is not based on scaling arguments that require many qubits and deep circuits but rather the ability to perform certain tasks, e.g., minimum error discrimination of quantum states, that are only possible with generalized measurements beyond the ones corresponding to simple intensity or heterodyne measurements of field modes. Moreover, some sensing/processing tasks tolerate errors, \emph{e.g.,} classification of images, and therefore approximate implementation of generalized measurements due to the errors in NISQ devices may be tolerable for such applications. 

\section{Example: joint detection receivers for coherent communication}
\label{sec:jdr}
Optical communication is the standard for today’s information networks. The increasing demand for bandwidth has led to the exploration of coherent transceivers that use phase- and amplitude-modulated optical signals to encode more bits of information per transmitted pulse. Although such coherent communication schemes can enable increased capacity communication, it is well understood that achieving the limits of communication bandwidth, especially in low light regimes, requires joint detection receivers (JDRs) that process multiple pulses coherently to decode \emph{codewords} instead of individually detecting and decoding one pulse at a time \cite{holevo_1998, Schumacher_Westmoreland_1997,Guha_2011}. The difficulty of designing and constructing such JDRs has prevented their adoption in the communications industry. We show that JDRs can be effectively constructed within the framework of QCIS and trained variational circuits.

In coherent communication, the transmitter encodes information in phase- and amplitude-modulated coherent pulses of light. Depending on the choice of encoding, one of a collection of $M$ coherent states of light are sent over the channel by the transmitter (thus each pulse encodes $\log_2(M)$ bits of information), and the task of the receiver is to discriminate between them to decide the identity of the transmitted pulse. Non-orthogonality of coherent states implies that the $M$ signaling states cannot be distinguished perfectly, and instead the optimal receiver performs this state discrimination with minimum average probability of error, or $p_{\rm err}$. Such optimal discrimination of non-orthogonal quantum states requires implementation of a generalized quantum measurement, and this is exactly the role of the JDR. 

Our approach to designing a quantum computer-enabled JDR proceeds via two stages: first, received optical pulses are transduced into states of superconducting qubits. We model this transduction with arguably the most mature optical-to-microwave transduction technology, optomechanical devices. Next, in stage 2, joint detection is executed by performing a quantum computation on the qubits with a shallow-depth, variational quantum circuit. The problem of designing a minimum-error receiver is then transferred to the problem of formulating a quantum circuit that optimally discriminates between the qubit states encoding the received pulses. 

\begin{figure}[t]
\centering
\includegraphics[width=0.5\columnwidth]{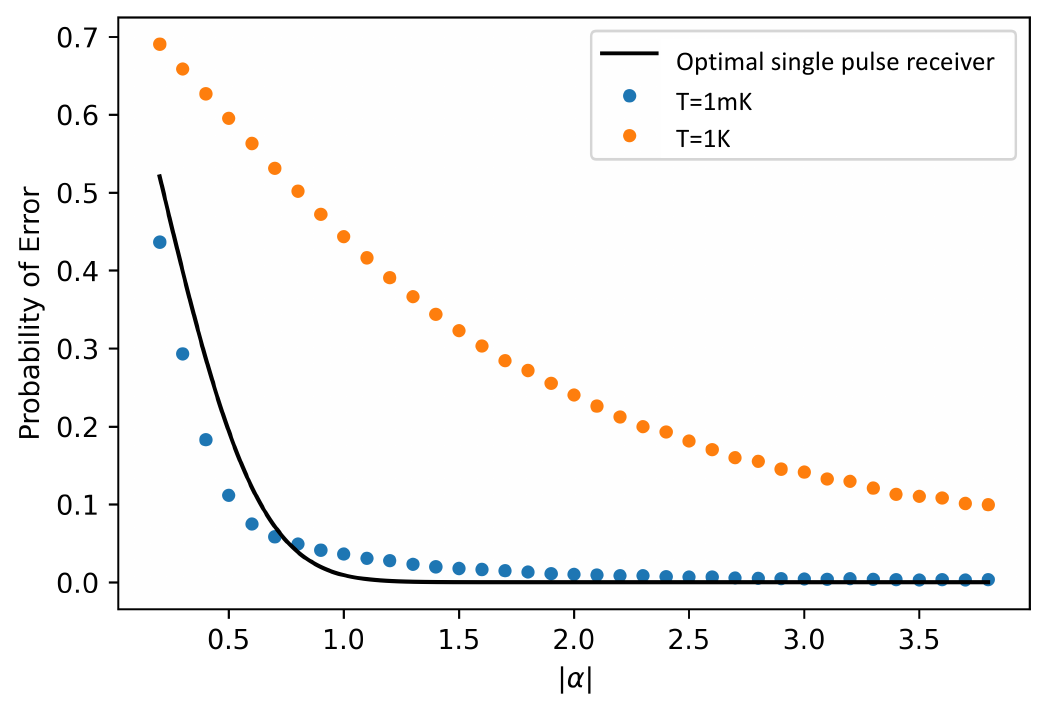}
\caption{$p_{\rm err}$ in decoding two bits encoded in three optical pulses using binary phase shift keying (BPSK), as a function of the magnitude of the received optical pulses. The black curve shows the optimal performance of the optimal ``classical'' single pulse receiver that operates in the optical domain (and thus decodes two pulses, which is sufficient to encode 2 bits using BPSK). The blue (orange) dots show the $p_{\rm err}$ achieved by transducing to qubit states and performing an optimized variational circuit to discriminate between the possible codeword states, when the optomechanical system performing the transduction is at 1mK (1K). \label{fig:jdr}}
\end{figure}

In Ref. \cite{Crossman_2023} we analyze in detail optomechanical transduction from optical to microwave and then into the state of qubits through a Jaynes-Cummings interaction between the microwave mode and qubit. This optomechanical transduction induces attenuation and heating of the received coherent states so that if the state $\vert \alpha e^{i\theta} \rangle$ is received, the transduced qubit state becomes a mixed state with the key phase information $\theta$ encoded in the azimuthal angle of the qubit's Bloch vector. In Ref. \cite{Crossman_2023} we outline the types of transduction fidelities to be expected for realistic optomechanical devices like the ones developed in Refs. \cite{Andrews_2014,Higginbotham_2018}, and perform training of variational circuits to distinguish the transduced qubit codewords. Typical results are shown in Fig. \ref{fig:jdr}, where we plot the achievable $p_{\rm err}$ for decoding 3-pulse codewords with the trained quantum circuits, for realistic values of transduction parameters and fidelities. This decoding error is compared to the best possible error probability achievable if decoding of individual pulses is done (we assume optimal decoding of each pulse, via an all-optical receiver that achieves the Helstrom bound \cite{Helstrom_1976}). As can be seen from this figure, depending on the transduction fidelity, which is largely determined by the transduction-induced heating and thus the temperature of the optomechanical system, and the received intensity, one can achieve $p_{\rm err}$ that is superior to the ``classical'' strategy of using single pulse decoding. This improvement in error probability is restricted to the very weak-field regime less than one photon received per pulse ($|\alpha|\ll 1$). This regime is relevant to deep space and satellite to earth communication channels. We conclude by noting that similar results were derived by Delaney \emph{et al.} using different models of transduction and computation \cite{PhysRevA.106.032613}.

\section{The path forward}
QCIS has the potential to deliver extraordinary gains in imaging and sensing precision, especially in the weak-field regime. As described in Sec. \ref{sec:jdr}, improvements over classical state-of-the-art are possible even with small (3-qubit) implementations of the idea with near-term hardware for quantum tranduction and computing. But to realize the spectacular quantum advantages envisioned by the complexity theory results in Refs. \cite{Aharonov_Cotler_Qi_2022,Chen_Cotler_Huang_Li_2021,Huang_2022}, there needs to be experimental and theoretical progress on several fronts, as we now briefly discuss.

Firstly, experimental progress in quantum transduction is required, to increase the fidelity of transduction of quantum states from the EM field to stationary qubits. Importantly, we need to think about transduction more broadly that in the quantum networking context, within which single excitation (single photon) transduction is the priority. For QCIS, it is beneficial to consider transduction of a broad set of properties of the EM field to qubits while preserving coherence and fidelity. This could require the development of new technologies for quantum transduction.

Secondly, more thought must be given to how transduction will integrate with scalable quantum computing architectures. This is a consideration for quantum networking, but is more important in the QCIS context where a large number of transduction channels must interface with qubits. 

Finally, on the theoretical front, more work is required to bridge the complexity theory results in Refs. \cite{Aharonov_Cotler_Qi_2022,Chen_Cotler_Huang_Li_2021,Huang_2022} and realistic imaging and sensing scenarios. Identifying real-world imaging or sensing applications that allow for the exponential resource separations derived in idealized settings in those works would provide significant impetus for the technology developments mentioned above.

\acknowledgments 
I would like to thank John Crossman, Spencer Dimitroff, John Kallaugher, Ashe Miller, Steve Young, Daniel Soh, and Lukasz Cincio for valuable discussions about the QCIS concept, joint detection receivers for classical communication and related ideas.
This material is based upon work supported by the U.S. Department of Energy, Office of Science, Office of Advanced Scientific Computing Research, under the EXPRESS program.
Sandia National Laboratories is a multimission laboratory managed and operated by National Technology \& Engineering Solutions of Sandia, LLC, a wholly owned subsidiary of Honeywell International Inc., for the U.S. Department of Energy's National Nuclear Security Administration under contract DE-NA0003525. 
This paper describes objective technical results and analysis. 
Any subjective views or opinions that might be expressed in the paper do not necessarily represent the views of the U.S. Department of Energy or the United States Government.

\bibliography{report} 
\bibliographystyle{spiebib} 

\end{document}